# Localization of states, Bloch wave and quantum phase transitions in SUSY deformed potentials in non-Hermitian optical systems

Muhammad Imran Afzal and Yong Tak Lee*

**M. I. Afzal** is affiliated with the School of Information and Communications, Gwangju Institute of Science and Technology, Gwangju, 500-712, Korea.

Telephone: 82-62-715-2239

Fax: 82-62-715-3128

Email: imran@gist.ac.kr

**Y. T. Lee** is affiliated with the School of Information and Communications, Department of Physics and Photon Science and the Advanced Photonics Research Institute, Gwangju Institute of Science and Technology, Gwangju 500-712, Korea.

Telephone: +82-62-715-2239

Fax: +82-62-715-3128

Email: ytlee@gist.ac.kr

Originally, Supersymmetry (SUSY) emerged as a theory of transformations between bosons and fermions, and became the strongest candidate to complete the standard model and thus establish the symmetry of the universe. However, the search for experimental verification has been problematic, and that has forced people to focus on experimentally realizable models to verify SUSY in limited systems. This has been experimentally adapted in Hermitian and non-Hermitian optical systems, however, has mainly been limited to the single-order transformation. In contrast to the single-order, the theory of higher-order transformations allows the deformation and redistribution of the energy of the original potentials by the annihilation of multiple groundstates.

For the first time, we have observed the annihilation of multiple eigenstates of the parent potentials and redistribution of the energy in the deformed potentials in the system with spontaneously breaking of parity-time symmetry while preserving the SUSY of the system. We have observed that the deformation of the potentials is sensitive to initial conditions. This first experimental observation enables the localization of eigenspectra in lateral space (so-called time) and the localization of eigenstates in longitudinal space in the deformed potentials by experimentally applying the SUSY higher-order transformations on tilted and disordered parent potentials, respectively. The former shows a decrease in the slope of the profile of eigenspectrum (bands) due to the localization in lateral space, while the latter shows an increase in the slope due to localization in longitudinal space. This allows the formation of deterministic bandgaps and Bloch waves while retaining the SUSY. In this configuration, the phase transition emerges naturally in the SUSY deformed potentials, which are revealed as the quantum Zeno effect and a peculiar Anderson localization, respectively. The Anderson localization shows the features of effective amplification and quantum Anti-Zeno effect. The experimental results are in full agreement with the theory of SUSY deformed potentials of higher-order transformations. Besides enhancing the understanding of nature, our results can also provide an experimental platform to originate new classes of effective and synthetic materials, and are also significantly important for unifying the interaction dynamics of condensed matter and photonic systems, and the implementation of universal quantum computation.

Understanding energy redistribution through dissipative interactions in multimode matter systems via non-equilibrium dynamics [1], and analogous studies in non-Hermitian [3] photonic systems via balancing gain and loss, could facilitate stronger tests of multimode interference, many-body entanglement, hidden variable theories, quantum gravitation and the implementation of universal quantum computation, to mention a few [1-6]. In both kinds of systems the exchange of energy could transform initial systems into different types of final systems depending on the initial conditions. For example, it is well established that the quantum Zeno effect, Anti-Zeno effect, nonclassical Anderson localization-like effect or effective amplification [7-9] are very sensitive to initial states. Investigating these requires a deep analysis of dissipative dynamics in nature, and may enable photonic and matter systems to be treated similarly. For example, the nonlinear transport of a many-body system in the absence of a macroscopic bias [10] (because of the asymmetry of space and the so-called symmetry of time or vice versa) is closely related to the quantum Zeno effect, which is caused by dissipation. This relationship could facilitate the cross-over of novel ideas and methodologies between apparently unrelated fields of science. An optical analogy of the resonant tunnelling of atoms of condensed matter systems can be used to condensate the photons in optical fibers [11].

Condensed matter interacting systems that are confined in the transversal dimension have emerged as promising candidates for the experimental study of many-body physics in nature [1]. These systems are also one of the most promising candidates for experimental implementation of universal quantum computation, particularly using the properties of quantum Zeno effect [8, 12], without the need for repeated initialisation of the system [7]. Intrinsically in these systems, the asymmetry of the space takes the form of an asymmetric trapping potential and the symmetry of time takes the form of a measurement. This set of features induces nonlinearity, such that the decay rate in the same system can be subsequently decelerated or accelerated simply by changing the dissipation rate. This behaviour is considered to be a most significant experimental manifestation of the quantum Zeno and anti-Zeno effects [12]. However, the dissipation can be controlled by slope of trapping potential while maintaining the balance between gain and loss. From this perspective, a new or deformed potential may be hypothesised to arise from the microscopic dissipative dynamics in a non-equilibrium condensed matter and a spontaneously breaking of PT-symmetric photonic system [13-15]. The continuous oscillation between different decay rates could cause the nonlinear transport of a wave packet without breaking [16], which would simulate Richard Feynman's ratchet-like effect. Equivalently, these dynamics could also be considered to be the oscillation between quantum Zeno and Anti-Zeno effects or equivalently, the oscillation of localization of energies between so-called time and space. Which are also experimentally accessible in photonic systems using modulated waveguides. In addition, photons exhibit more fundamental dynamics than the material particles of condensed matter systems. This feature, along with the ease of analogical experiments has enabled researchers to confirm a variety of theoretical and experimental features of non-photonic origin and provided a novel means of extending our understanding of quantum mechanical laws over the last few decades [17].

Numerous efforts have been aimed at dynamically generating a condensed matter-like interaction of pure photons via dissipation and sensitive to initial states beyond the few photons dissipative interaction. The most significant approach is the present front-runner, superconducting circuits, which are applicable to larger numbers of photons [18]. Remarkably, in the context of superconducting circuits, a Josephson junction can self-drive the system towards quantum coherence and phase-shift at a macro level through dissipation. Consequently, the system can self-transit from weak to strong coupling, producing several quantum mechanical phenomena [19-20]. However, there are major hurdles to achieving condensed matter-like dynamics in optical systems, including the photon coupling loss between wave packets and the superconducting circuit, and leakage errors, which still hampers the self-transitions from parent systems to new systems i.e. deterministically higher-order deformations of parent systems. The ideal solution is, that a simple photonic system itself perform dissipative oscillation or oscillation between balanced the gain and loss. In this context implementing SUSY higher-order transformations [28] in a PT-symmetric [3] system by implementing the balance between gain and loss using resonant interference [32] and spontaneously breaking of PT-symmetry [24] can provide the promising solution instead of using gain medium and lossy waveguides.

SUSY was originally proposed a long time ago to describe the symmetry of the universe. The elegant theory of SUSY provides many intriguing deep features which are not allowed in other competitive theories. This makes SUSY the only candidate to complete the standard model [21], by providing transformation between bosons and fermions, however, experimental verification is still lacking. The long wait for experimental verification has given birth to many controversies about supersymmetry, and because of the experimental difficulties involved, many groups and individuals are now focussing on an experimentally realizable SUSY models in limited systems. One of these efforts is to find SUSY in the topology of phase transitions in a condensed matter system [22]. Another important effort is the investigation of SUSY non-relativistic quantum mechanics [23].

The theory allows the formation of new structures while keeping a definite relation with original structures through universal phase matching and redistribution of the energy of annihilated eigenstates. The real beauty of SUSY comes into play when higher-order transformations are applied, which allows the formation of complex structures. These are the prominent features which SUSY quantum mechanics shares with the theory of SUSY in particle physics [21]. These concepts have been successfully applied in optics, on propagating waves, to produce universal phase-matching and isospectral characteristics [13, 24]. They have been experimentally demonstrated in a number of systems made of coupled waveguides [25-26] and photonic lattices [27].

Progressing in this direction, it was recently, theoretically proposed that deformed potentials by SUSY higher-order transformations can produce tunable bandgaps, localization, and Bloch wave [28], however, additionally, we noted that the higher-order transformations of the potentials naturally exhibits quantum phase transitions. In another theoretical study, the self-transition of the phase is also found by apply SUSY transformations on topological condensed matter system [22].

Optical lattices and arrayed waveguides are widely used for engineering desired potentials to explore the interaction dynamics of condensed matter and photonics respectively. Particularly, engineered potentials are very useful for the study of quantum phase transitions and entanglement. In particular disordered and tilted (washboard) potentials are widely used to produce the Anderson localization [29], quantum Zeno and Anti-Zeno effects [12]. In particular the quantum Zeno effect [7] intrinsically contains the (nonlinear) quantum mechanical features and dynamics, which can transform a multimode classical system into a pure quantum mechanical system. This effect has no analogue. The quantum Zeno effect can also be used to explore the innovated and deterministic methods of quantum computation [8], thermodynamic quantum cooling [30] and implementations of Demon-like algorithms [31].

In the present study we used tilted and disordered potentials as parent potentials and experimentally applied SUSY higher-order transformations to produce the localization of the eigenspectrum in time and space respectively. A detailed drawing is shown in Fig. 1., summarizing the concept and important features of the experimental results. In our experiment the two parent systems were made of tilted and disordered potentials by nine eigenstates and five eigenstates, respectively, as shown in Fig. 3(a) and 4 (a) respectively.

The tilted parent potential contains nine eigenstates of effective refractive indices, and the order is described as:

$$(n_1, n_2, n_3 \ldots n_9) = \begin{pmatrix} n_1, n_1 + \Delta n, n_1 + 2\Delta n, n_1 + 5\Delta n, n_1 + 6\Delta n, \\ n_1 + 7\Delta n, n_1 + 7\Delta n, n_1 + 10\Delta n, n_1 + 11\Delta n \end{pmatrix} \quad \ldots(1)$$

These nine effective indices (eigenstates) functioned as nine (effective) waveguides. However, $n_6$ and $n_7$ have the same effective index, so effectively the system of the parent potential contains eight eigenstates. The disordered parent potential has five eigenstates, and the order is represented as:

$$(n_1, n_2, n_3, n_4, n_5) = (n_1, n_1 + 4\Delta n, n_1 + 5\Delta n, n_1 + 7\Delta n, n_1 + 8\Delta n) \quad \ldots(2)$$

The wave function ($\psi$) of the parent tilted (disorder) potential is described by using Wannier states as

$$\psi = \cos(2\pi r + \phi)|n_r\rangle\langle n_r| \quad \text{...........................................................................................(3)}$$

$\phi$ is an arbitrary phase of tilted (disordered) potential.

The parent potentials contained eigenstates in lateral space and longitudinal space, which symbolizes the states in time and space as depicted in Fig. 3 (a) and Fig. 4(a). In Fig. 1 (c, f) the mesh of black and green lines are drawn to visualize time and space respectively, and the positions of the occupied states in the parent potentials in lateral space are represented with $n_r$. The location of the orders of the nine and five eigenstates in time in tilted and disordered parent potentials are represented by the Eq. (1) and Eq. (2) respectively. Meanwhile the eigenstates in the space are distributed uniformly with longitudinal spacing of 0.54 and 0.63 nm. The presence of the eigenstates in respective parent potentials in such orders makes it possible to visualize the tilted (disordered) parent potential system, analogous to the famous tilted (disordered) potentials produced by Wannier states (standing waves) in condensed matter systems [12, 29].

To experimentally apply SUSY higher-order transformations on the parent potentials (tilted and disordered), we used the experimental setup mentioned in Fig. 2. We resonantly interfered the tilted (disordered) potentials at a 50:50 coupler. The gain and loss was naturally balanced because the resonant interference only allows the tunnelling of resonant photons from waveguide 1 to 2 and vice versa. This configuration allows dynamic resonant tunnelling for an instant and then reinstates inhibition of the tunnelling between the waveguides. Therefore, the role of the coupler in our experiment is limited to the initialization of tunnelling among the adjacent effective waveguides, formed by the tilted (disordered) potentials in the same single waveguide.

The tunnelling inhibition occurs because tunnelling of photons from waveguide 1 to 2 induces tunnelling in adjacent modes, and this latter tunnelling effect make the systems in waveguide 1 and 2 nonresonant, and thus independent from each other. Thus the PT-symmetry between adjacent modes is maintained and after certain threshold the localization of states and phase transitions occurred. It is also important to note that theoretically, it well-known that PT-symmetric systems [3] are sensitive to initial conditions, particularly, this sensitivity is pronounced when the (dissipative) oscillation between exactly balanced gain and loss continue for sufficient long time, which ultimately breaks the PT-symmetry. In this configuration the system can exhibit SUSY higher-order transformations as well as quantum phase transitions [24, 32].

The tunnelling between adjacent modes in the same waveguide is completely analogous to the tunnelling in condensed matter systems formed by trapping cold atoms in standing waves [12, 29, 33]. In quantum optics a recent and an important relevant experimental effort is the Demon-like cooling of optical modes [31]. However, contrary to condensed matter systems, we can clearly observe the annihilation of eigenstates and redistribution of their energies in the optical system. This aspect of SUSY in optics enables us to observe the SUSY, Bloch wave and quantum phase transition in a single experiment, a feat, which is inaccessible in condensed matter systems. In addition, the photonic systems are highly stable and longer coherence length compared to their condensed matter counterparts.

In this configuration, the nine (five) eigenstates in the tilted (disordered) potential function like a lattice having nine (five) effective waveguides whose wavefunctions are overlapped. This allows us to experimentally achieve the SUSY higher-order transformations of tilted and disordered parent potentials. Ultimately the eigenspectrums are localized in time and space respectively. Theoretically the index profile $n_s\left(\overset{\cdot}{x}\right)$ of SUSY deformed potentials can be described by the equation [28]:

$$\left[n_s\left(\dot{x}\right)\right]^2 = \left[n_r\left(\dot{x}\right)\right]^2 + 2\frac{1}{k_o^2}\frac{d^2}{dx^2}(\log\psi) \quad \text{...............................................................(5)}$$

where $\left[n_r\left(\dot{x}\right)\right]^2$ is tilted (disordered) potential and $\left[n_s\left(\dot{x}\right)\right]^2$ is the corresponding higher-order SUSY deformed potential. $k_o$ is the free-space wavevector.

However, it is important to note the limitation of SUSY quantum mechanics [23], because this only supports SUSY in time only. As the experimental studies are free from these kinds of limitations of the theories therefore we observed the localization of original eigenspectrum in space for the disordered potential. Consequently, Anderson localization is emerged in time as shown in Fig. 4(b). Nevertheless, we envisage that by applying the theory of Space-Time SUSY of particle physics and topology [22] on our experimental results, it will unfold more interesting details.

Owing to SUSY higher-order transformations, five and two groundstates of the tilted and disordered potentials are annihilated while whole of the eigenspectrum of the parent potentials are retained in respective deformed potentials, which are observed as localization in time and space respectively. While the relative spacing of the eigenstates both in time and space is preserved in the SUSY deformed potentials, as shown in Fig. 3., and Fig. 4., by capturing the parent and SUSY potentials of both systems. The redistribution of the energies formed the bandgaps and Bloch waves in time (lateral space), while preserving the eigenspectrum of parent potentials, which is clearly visible in Fig. 3(d) and Fig. 4(d). In the SUSY deformed potentials, the localization of the eigenspectrum of the tilted (disordered) parent potential causes a decrease (increase) in slope of the profile of eigenspectrum as shown with straight and thick blue and red lines in Fig. 3(a, b) and Fig. 4(a, b).

This character shows natural emergence of the quantum phase transformation at macro scale, which is evident with the observation of quantum Zeno effect (Anderson localization). Where the quantum Zeno effect is naturally emerged due to the localization of the eigenspectrum in time and Anderson localization is emerged due to the localization of eigenspectrum in the space. In Fig. 4(b), apparently, the SUSY deformed potential has eigenspectrum within the region of annihilated eigenstates ($n_{4-5}$) in the time domain, which is not allowed by SUSY transformations. But actually, the preservation of the power (0.038) of central mode and the order of eigenstates shows that the whole deformed potential is shifted upward to preserve the SUSY. Ultimately, the disordered parent potential is preserved in the deformed potential and is localized in space instead of time domain. In such a way the SUSY is preserved in the deformed potential. However, it seems that that eigenspectrum of two lowest eigenstates ($n_1$ and $n_2$) in the deformed potential is localized in time.

This observation allow us to introduce an interesting postulate: If a SUSY deformed potential could not preserve the eigenspectrum of the parent potential in time (or in a specific dimension) then it will localize the eigenspectrum of parent potential to other dimension to obey the SUSY. This aspect is important to study the increase of the effective dimensions in any local system and thus growth of a system. This is interesting for further investigation of interaction of eigenstates which are divided between more than one dimension and to study the increase of the effective dimensions in any local system.

Our observed Anderson localization is very different from the conventional Anderson localization in optics, however, it is in agreement with the Anderson localization [29] and Anti-Zeno effect [12] in condensed matter and effective amplification [9] in quantum optics, The effective amplification [9] is reminiscent of quantum optics and a necessary element for experimental implementation of quantum computation, quantum detection and Schrodinger's cat based amplification of the quantum states. In particular, comparing Fig. 4 (a) and Fig. 4(b) in terms of effective amplification of modes instead of redistribution of energies of parent eigenstates, in that aspect, we observe that the spacing between the central mode and sideband increased twofold from 0.54 nm to 1.08 nm, accurately 0.63 to 1.26 nm, The mode at $\simeq 1007.86$ nm is amplified from $\simeq 0.103$ to $\simeq 0.128$, and the effective amplification ranges from −24% (as indicated by the blue downward arrow) to 62% and 92% (as indicated by the red upward arrows). It is extremely important to note that the relative power (0.038) of the signal at $\simeq 1007.86$ nm is almost the same for the parent and the transformed system. Preservation of the power or energy along with shape of the mode is very important in SUSY transformations, because this kind of preserved mode always replaced the first annihilated groundstate in SUSY deformed potentials. Particularly, the feature of preservation of power and shape of the peak of the mode (which contains first groundstate in parent potentials) in deformed potential is observable in Fig. 4(b) and also in Fig. 3(b).

These peculiar localizations, which have no other equivalents in optics, also allowed us to observe the reduced intensity fluctuations of the systems as shown in Fig. 5., in particular, redistributed energies in the quantum Zeno effect profile (Fig. (a) shows antibunching and clustering-states like effects).

In summary, the modes of parent potentials resonantly interfered at the coupler and initially gain and loss at the coupler triggered the interaction between adjacent modes (effective waveguides) and then inhibited. Subsequently, the dissipative oscillation between adjacent modes is continued for sufficient long time and spontaneously breaks the PT-symmetry [24], which ultimately produced SUSY higher-order deformed potentials. Our experimental results demonstrated that the SUSY higher-order deformation of tilted and disordered potentials not only provide bandgaps and Bloch-waves without using Bloch's theorem [34] but also naturally produce the quantum Zeno effect and Anderson localization in the system as a result of spontaneously PT-symmetry breaking. Consequently, the system exhibited quantum Zeno effect and Anderson localization depending on initial state of the parent potential. Formation of bandgaps and Bloch's wave are in agreement with a very recently reported theoretical study [28]. These observations provide the first experimental verification of SUSY higher-order transformations of parent potentials. The order of the eigenstates in the parent potential controls the energy of the bands in the SUSY deformed potentials, and largest gap(s) of eigenstates formed the bandgaps, which can be described using the theory presented in [28]. At the same time, the tilt and disorder control the slope of the bands distribution in the deformed potentials, or simply, the quantum phase transition. This feature is in agreement with the quantum phase transition in condensed matter systems [12, 29].

Therefore, the localization of eigenspectrum of the parent potential in time produced the quantum Zeno effect, while the localization in space produced peculiar Anderson localization in time, showing features of effective amplification and quantum Anti-Zeno effect. The results are significantly important for understanding the deeper physics of nature, in particular the nature of eigenvalues in deformed potentials [35-37]. Some of the immediate effects are expected to be in the fields of tunable bandgap engineering, originating new classes of effective and synthetic materials, condensation of photons, Mott insulation and superfluidity [38], highly stable microlasers [39] unifying the dynamics of condensed matter and photonic systems, topology, quantum cooling, enhancing sensitivity of photodetection via the effective amplification, and generation of deterministic cluster-states [40] or the photonic bands for implementation of universal quantum computation and dissipation based quantum simulator [4-5]. In particular, we envision that further advanced experimentation and explanation of results using topology will enable stronger visualization of the dynamics of directly inaccessible systems such as biological systems, quantum optical systems, very high and very low energy system, changes at the edge of system's state and phase transitions etc.

## Acknowledgements

This work was partially supported by the "Systems biology infrastructure establishment grant" and "Asian Laser Center program" provided by GIST.

## Author Contributions

M. I. Afzal conceived the idea, designed and performed the experiments and carried out analysis. M. I. Afzal and Y. T. Lee wrote the manuscript. Y. T. Lee supervised the project.

**Competing Interests Statement**

The authors declare that they have no competing financial interests.

 **Figure Legends**

**Figure 1 | Schematic of tilted (disordered) parent potential and localization in time and space in SUSY deformed respective potentials a, d,** show the distribution of eigenstates in two dimensions forming the geometry of tilted (disordered) parent potential. **c, e,** the distribution of eigenstates is more clearly imaginable by depicting the eigenstates in a mesh of eigenstates in time (black horizontal lines) and space (vertical green lines), where $n_r$ represent the eigenstates which are present in tilted (c) and disordered (e) potential. Numbers 1, 2, 3 in (a) and 1, 2 (d) shows virtual bands. Blue tilted lines represent the slope of the parent potentials. **b, e,** show the localization of the eigenspectrum or effective localization of eigenstates in lateral and longitudinal space: this localization form the bandgaps with three bands (1, 2, 3), two bands (1,2) and Bloch waves after SUSY deformation of the tilted (disordered) potential a(d). The bands are shown with red histograms.

**Figure 2 | Experimental configuration:** The initial tilted (disordered) potential is injected into a two-metre loop through the dense wavelength division multiplexer (DWDM). The entire loop consists of a normal dispersion optical fibre (HI 980). The fibre exhibits zero dispersion at $\simeq 1300$ nm, and the dispersion values at 980 nm are $\simeq -63$ ps/nm/km. The normal dispersion fibre is used to prevent any unexpected effects from the intrinsic instability of the material. The eigenstates in the input potential self-interfere resonantly at the coupler, which causes the balance of loss and gain for an instant only and then the interference becomes nonresonant. This self-generated nonresonance at the coupler between waveguide 1 and 2 causes inhibition of tunnelling between the two waveguides, however resonant tunnelling among adjacent states continues which ultimately spontaneously breaks the PT-symmetry which enables the operation of SUSY higher-order transformation. Consequently, 1) tilted (disordered) parent potential is deformed, 2) the bands are formed, 3) localization of the eigenspectrum in time (space) occurs. The experiments were performed at very low power (the peak power of the wave at the detector is 3.38 μW after passing through the experimental apparatus) as shown in Fig. 3(a), and the frequency spacing between the consecutive stationary modes (eigenstates) is sufficiently long (0.54, 0.63 nm). An optical

spectrum analyser (OSA) (Anritsu MS9710 C) with a sensitivity of -85 dB has proved to be sufficient to measure the changes in the spectral properties of the travelling wave.

**Figure 3 | Experimental generation of SUSY deformed potential, annihilation of five groundstates, localization in lateral space (so-called time), Bloch wave formation and quantum Zeno effect, a, c** A photonic wave packet exhibiting washboard-like potential, where the Wannier states-like modes have a constant spacing of 0.54 and 0.63 nm, a peak wavelength of 1000.66 nm and a power of 3.38μW , is detected at the output of the experimental apparatus just before the SUSY transformations occurred. The blue straight line shows the slope of the distribution of the spectrum of eigenstates in the parent potential **b, d** Annihilation of five eigenstates ( $n_{-5\ 9}$ ) of the parent potential and generation of deformed potential after SUSY higher-order transformations. Longitudinal and lateral space symmetries are preserved and localization happened in lateral space. However, PT-symmetry is spontaneously broken. The peak power is reduced to 2.04 μW and the energies of the annihilated states are redistributed while retaining the SUSY. The red straight line shows the increase in slope of redistributed eigenspectrum which are localized in literal space. The change in slope is interpreted as the quantum Zeno effect. In the inset: $g_e$ and $g_z$ show the exponential slope of the parent potential and the slope of the SUSY transformed (quantum Zeno effect) potential. **d** Red histograms encircled with black rectangular shapes show bandgaps: their periodicity forms the Bloch wave.

**Figure 4 | Experimental generation of SUSY deformed potential, annihilation of two groundstates, localization in longitudinal space, Bloch wave formation and Anderson localization. a** A parent potential having disordered eigenstates with consecutive gaps in longitudinal space is 0.54 and 0.63 nm. The potential (wave packet) is detected at the output of the experimental apparatus just before the resonant interference occurs. **b** SUSY deformed potential exhibiting localization of the parent eigenspectrum in the longitudinal space, which is evident with the increase in the slope (red straight line) exhibiting features of Anderson localization, effective amplification and Anti-Zeno effect. However, it is important to note that

eigenspectrum of two lowest eigenstates ($n_1$ and $n_2$) localized in time. The blue and red arrows indicate the power ratio of the mode at ~1007.86 nm relative to the sidebands before and after SUSY transformations. Note that the absolute power (0.038) of the mode is almost the same in the parent and SUSY potential. The spacing between consecutive modes increases two-fold from 0.54(0.63) nm (see in **a**) to 1.08 nm, and almost three-fold suppression of power of the sidebands. There is no effect on the boundary mode during the transformation from **a** to **b**. **d** Bands are represented with encircled red histograms and thus the Bloch wave.

**Figure 5 | Comparison of the ascending power distribution in tilted (Fig. 3(a)) and disordered (Fig. 4(a)) parent potential and SUSY deformed Fig. 3(b) and 4(b) respectively. a** The power for an arbitrary number of photons in the tilted potential is shown in **blue**, and the power redistribution in SUSY deformed potential is shown in **red**. **b** The power for an arbitrary number of photons in disordered potential is shown in **blue**, and the power redistribution in SUSY deformed potential is shown in **red**.

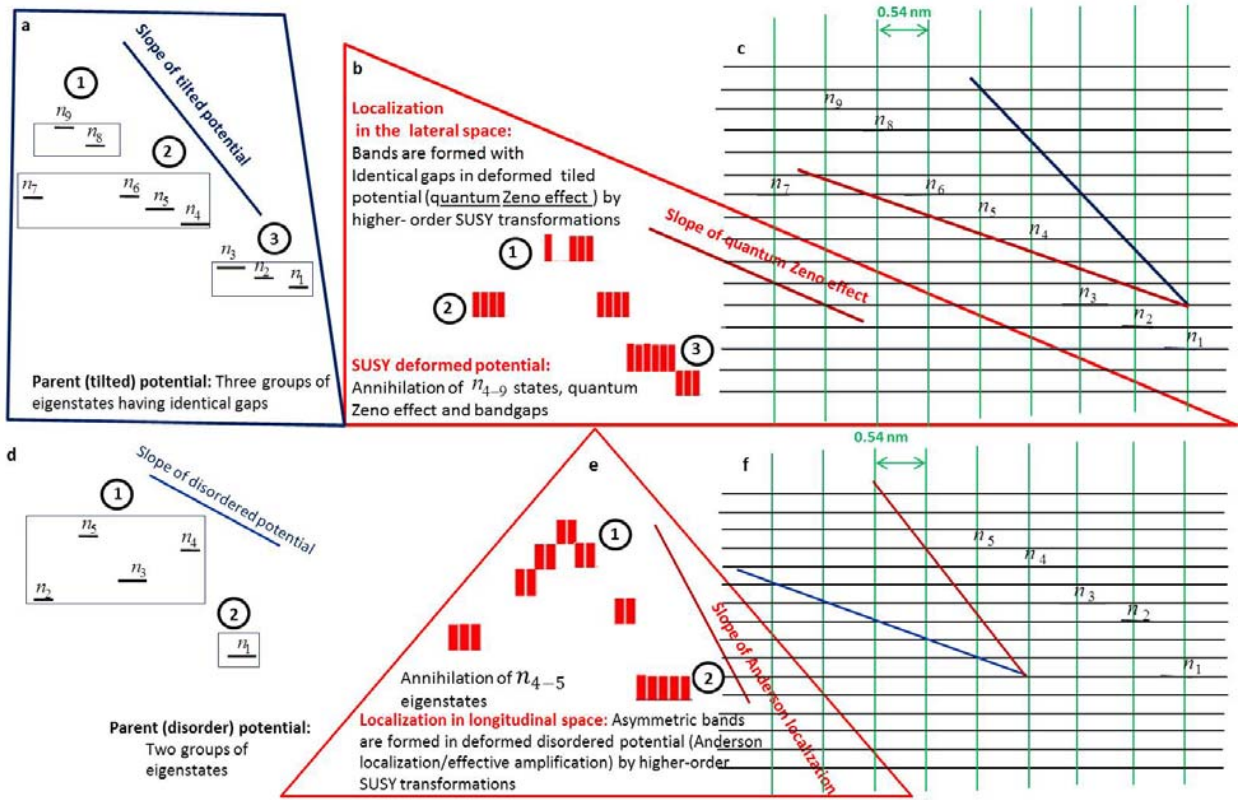

**Figure 1**

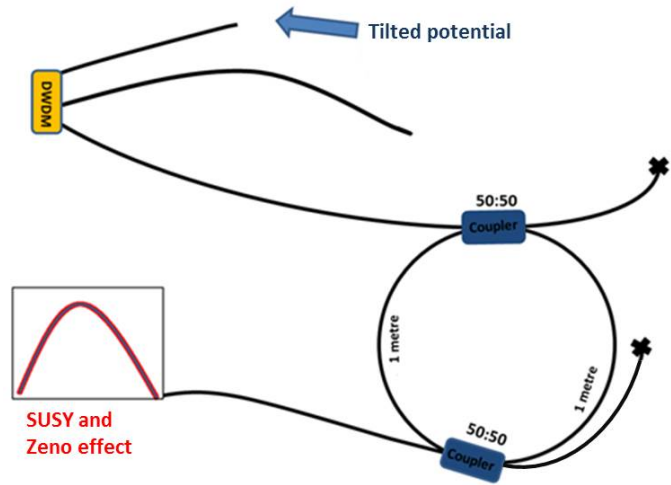

**Figure 2**

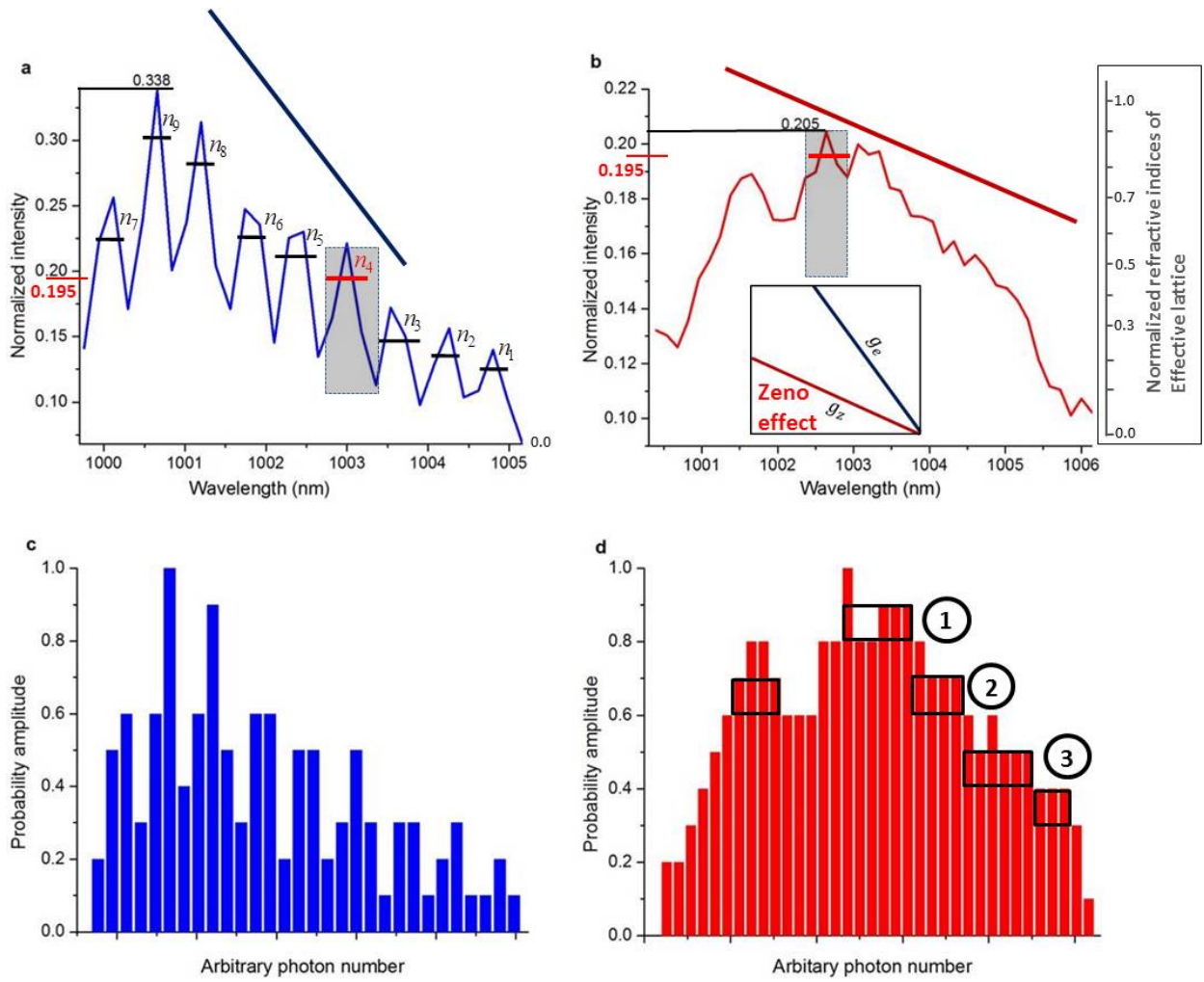

**Figure 3**

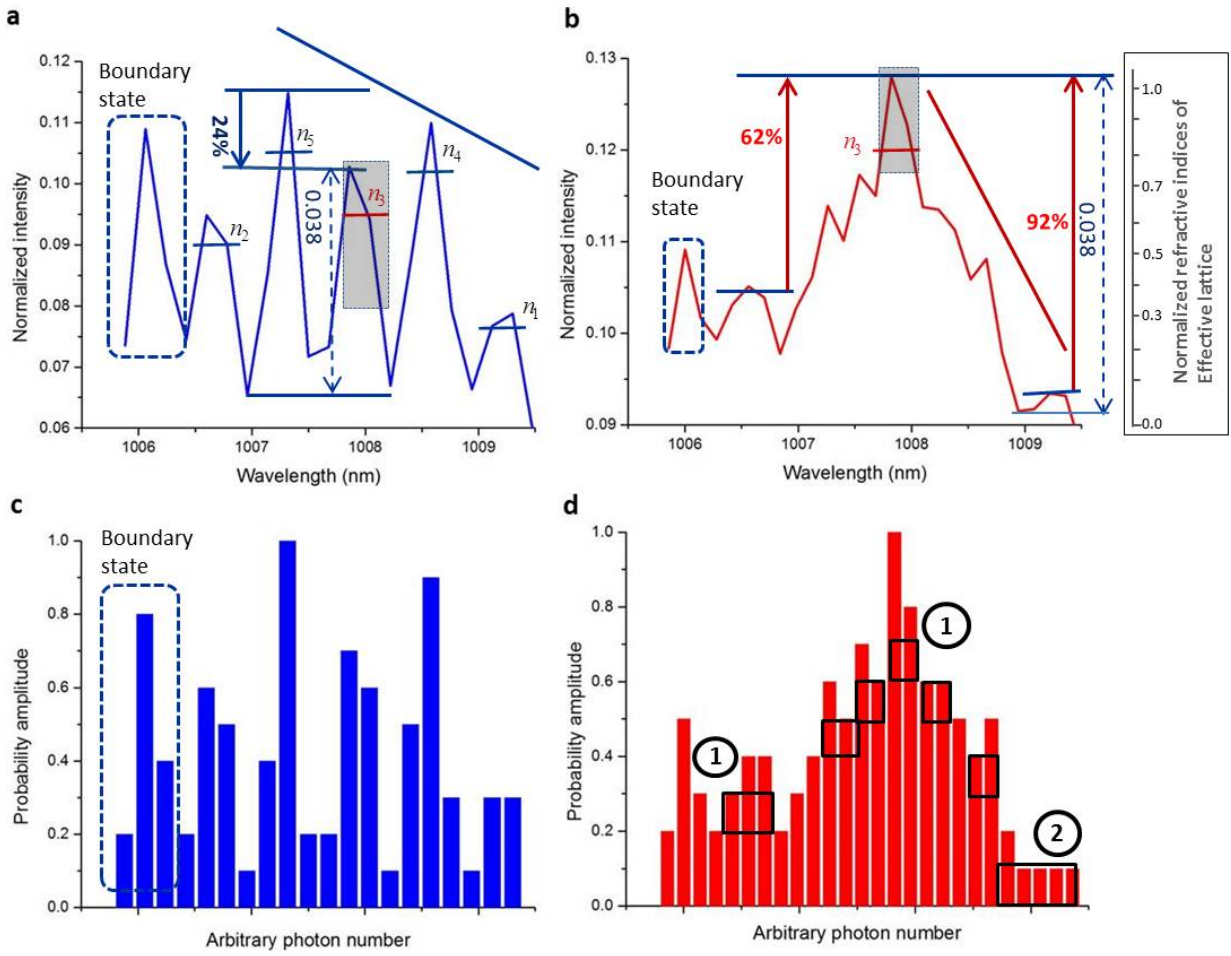

**Figure 4**

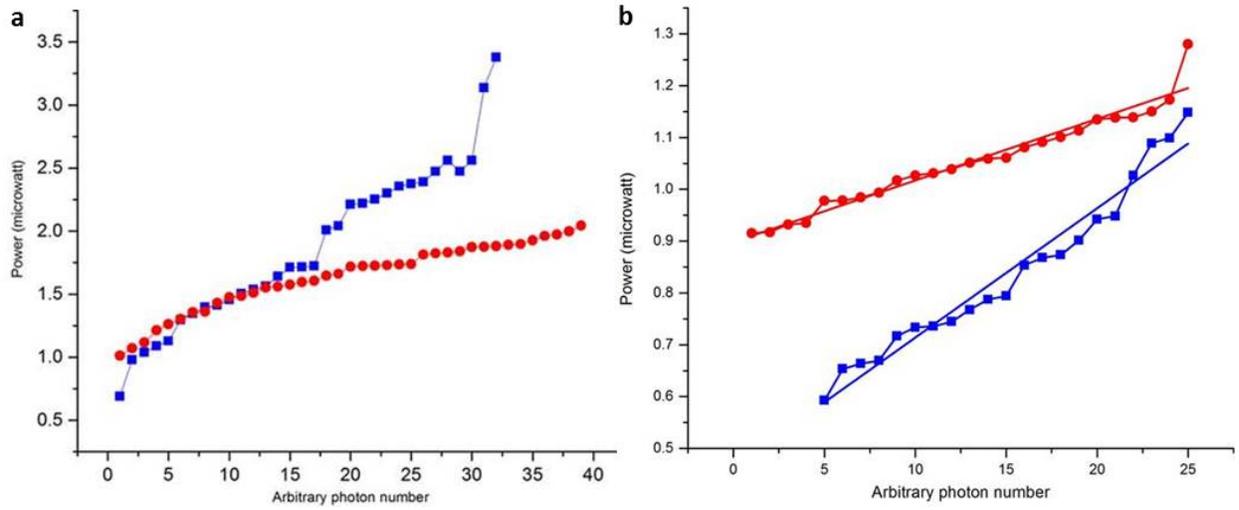

**Figure 5**